\documentclass[12pt]{article}
\usepackage{latexsym}
\begin{document}
\rightline{IU-MSTP/72}
\rightline{August, 2005}

\vspace{5mm}

\begin{center}
\LARGE{Operator ordering and Classical soliton path\\
in Two-dimensional $N=2$ supersymmetry\\
with K\"{a}hler potential}

\vspace{12mm}
\def\thefootnote{\fnsymbol{footnote}}
\large{Nobuyuki MOTOYUI\footnote[1]{motoyui@serra.sci.ibaraki.ac.jp} and
Mitsuru YAMADA\footnote[2]{m.yamada@mx.ibaraki.ac.jp}}
\def\thefootnote{\arabic{footnote}}

\vspace{12mm}
\large{Department of Mathematical Sciences, Faculty of Sciences,\\
Ibaraki University, Bunkyo 2-1-1, Mito, 310-8512, Japan}

\end{center}

\vspace{12mm}

\begin{abstract}
We investigate a two-dimensional $N=2$ supersymmetric model which consists
of $n$ chiral superfields with K\"{a}hler potential.
When we define quantum observables, we are always plagued by operator
ordering problem. Among various ways to fix the operator order, we rely
upon the supersymmetry. We demonstrate that the correct operator order is
given by requiring the super-Poincar\'{e} algebra by carrying out the
canonical Dirac bracket quantization. This is shown to be also true when
the supersymmetry algebra has a central extension by the presence of
topological soliton.
It is also shown that the path of soliton is a straight line in the
complex plane of superpotential $W$ and triangular mass inequality holds.
One half of supersymmetry is broken by the presence of soliton.
\end{abstract}

Keywords: Supersymmetry; operator ordering; soliton; Bogomol'nyi bound.

\newpage
\section{Introduction}

In the process of quantization of field theory with curved target space,
e.g. nonlinear sigma model, we are usually plagued by the operator
ordering in the definition of various quantum observables. We must find
out properly ordered quantum operator $\hat{F}$ from classical dynamical
variable $F(x,p)$. Various traditional ways are known for this problem.

The first way of them comes from path integral. The midpoint prescription
in the path integral leads to the Weyl ordered operators:
\begin{eqnarray}
\int \frac{dp}{2\pi \hbar} e^{-ip(x-x')} H(\frac{x+x'}{2},p)
= \langle x'| H(x,p)^{Weyl} |x\rangle
  \label{eq:int-0a}
\end{eqnarray}
where $H(x,p)$ is any classical dynamical variable \cite{das}.
However the Weyl ordering does not always give sufficient answer, as the
following examples indicate.

The second approach is connected with the self-adjoint extension of
Hermitian operators. To illustrate it, let us pick up a Hermitian operator
\begin{eqnarray}
F = x^{3}p + px^{3}.
  \label{eq:int-0b}
\end{eqnarray}
This has a pure imaginary eigenvalue
\begin{eqnarray}
F\psi = -i\psi
  \label{eq:int-0c}
\end{eqnarray}
with eigenfunction
\begin{eqnarray}
\psi_{\pm} (x) = |x|^{-\frac{3}{2}}e^{-\frac{1}{4x^{2}}} \theta (\pm x)
  \label{eq:int-0d}
\end{eqnarray}
where $\theta (x)$ is a step function.
On the other hand $+i$ is not an eigenvalue. Therefore the index of this
operator is (0,2) which states that $F$ does not have a self-adjoint
extension. Therefore an appropriate ordering of $2x^{3}p$ should be
sought out. However the Weyl ordering makes no use here because
$F' = x^{2}px + xpx^{2} = F$.
Therefore some other prescription is needed to obtain the quantum
counterpart of $2x^{3}p$ \cite{rud}.

The third approach is connected with symmetry: how symmetry dictates the
operator ordering. Mostafazadeh fixes the order of operators which
appear in supercharge $Q$ by Peierls bracket quantization
in supersymmetric system \cite{mosta}.
Pursuing this approach, we consider a two-dimensional $N=2$ Wess-Zumino
type model which consists
of $n$ chiral superfields with K\"{a}hler potential $K(\phi ,\phi^{\ast})$.
When this term is flat,
$K(\phi ,\phi^{\ast})=\phi\phi^{\ast}$, there was no ordering problem
because the target space is flat \cite{mty}.
We shall admit that general K\"{a}hler potential raises the ordering
problem. We rely upon the supersymmetry algebra \cite{soh,wb}
to fix the operator ordering: we will fix the operator orders by requiring
the super-Poincar\'{e} algebra.

When the kinetic term has a nonflat K\"{a}hler potential,
$K(\phi ,\phi^{\ast}) \neq \phi\phi^{\ast}$, the operator ordering problem
appears. There are several nonequal operator orders
in the supercurrent and canonical momentum operator.
To fix the operator orders correctly,
we require that each component fields $\varphi$ satisfy
the following relations:
\begin{eqnarray}
-i\lbrack \varphi, Q \rbrack_{\pm}
-i\lbrack \varphi, \bar{Q} \rbrack_{\pm} &=&
\delta\varphi
  \label{eq:int-1}
\end{eqnarray}
where $Q$ and $\bar{Q}$ are supercharges. When $Q$ and $\bar{Q}$ satisfy
this relation we can obtain the correct supersymmetry algebra.
Then we can take it as the correct operator order. This is also true when
the supersymmetry algebra has a central extension by the presence of
topological soliton.

In the supersymmetric field theory the properties of soliton solution is
studied long ago. As Witten and Olive pointed out, in the two-dimensional
$N=1$ supersymmetric theory the supersymmetry algebra is modified to 
include central charges in the presence of topological soliton \cite{wo}.
The phenomenon which breaks a part of supersymmetry is studied since
1980's \cite{bw}. It also happen in our model in the presence of
soliton which saturates the Bogomol'nyi mass bound.
In the presence of soliton, there are central extensions in
the two-dimensional $N=1$ supersymmetry \cite{wo}
and the two-dimensional $N=2$ supersymmetry with a flat metric \cite{mty}.
We can obtain a central extension in the two-dimensional $N=2$ supersymmetry
with nonflat K\"{a}hler metric in the presence of soliton.
In general soliton is a curve which connect each zero energy solution.
On the other hand we can obtain a straight line
in the complex plane of superpotential.
Then we can obtain triangular inequality
\begin{eqnarray}
M_{IK} < M_{IJ}+M_{JK}
  \label{eq:int-2}
\end{eqnarray}
among the classical mass of solitons.
It means that there is attractive force between neighboring solitons.
There isn't marginal stability such as in refs. 9 and 10
because the equality
does not hold. Because the classical mass of soliton saturates
the Bogomol'nyi mass bound, it breaks a half of supersymmetry.

This paper is constructed as follows. In Sec. 2 we construct
the two-dimensional $N=2$ supersymmetry which consists of two-dimensional
chiral superfield $\phi$. We derive canonical quantization conditions
through Dirac brackets. In Sec. 3 we fix the operator orders in $j^{\mu}$
and $\pi_{a^{i}}$. We obtain the correct operator order in $j^{\mu}$
and $\pi_{a^{i}}$ which satisfy the correct supersymmetry algebra.
It contains a central charge. In Sec. 4 we apply Hamilton-Jacobi method of
classical mechanics to bosonic Lagrangian. We obtain a straight line
of soliton path in complex $W$-plane. Classical mass of soliton saturates
Bogomol'nyi bound and it satisfy triangular mass inequality.
In Sec. 5 we show the multiplet shortening of supersymmetry algebra.
In Sec. 6 we check the ``central charge'' $T$ commutes with other operators.
Section 7 is devoted to conclusion of the work.
\section{Noether current and Dirac bracket quantization}

We consider the two-dimensional $N=2$ supersymmetric theory. A chiral
superfield is given by
\begin{eqnarray}
\phi &=&
a(x_{+}) +\sqrt{2}\bar{\theta}^{c}\xi(x_{+}) 
+\bar{\theta}^{c}\theta f(x_{+})
  \label{eq:ord-1}
\end{eqnarray}
where $x^{\mu}_{+}=x^{\mu} + i\bar{\theta}\gamma^{\mu}\theta$ and
$\xi$ and $\theta$ are two-dimensional Dirac spinors. We use
2-dimensional $\gamma$ matrices in the following representation:
\begin{eqnarray}
\gamma^{0}~=~\sigma_{2},~~
\gamma^{1}~=~-i\sigma_{1},~~
\gamma_{5}~=~\gamma^{0}\gamma^{1}~=~-\sigma_{3}.
  \label{eq:ord-1a}
\end{eqnarray}
When the kinetic term has a nonflat K\"{a}hler potential, the Lagrangian
of Wess-Zumino type model of $n$ chiral superfields $\phi^{i}$
is given as
\begin{eqnarray}
{\cal L}&=&
\int d^{2}\theta d^{2}\theta^{\ast} K(\phi^{i},\phi^{\ast i})
+\int d^{2}\theta W(\phi^{i})
+\int d^{2}\theta^{\ast} W(\phi^{\ast i}) \nonumber\\
&=&
\partial_{\mu}a^{\ast i}K_{i^{\ast}j}\partial^{\mu}a^{j}
+\frac{1}{2}iK_{i^{\ast}j}\bar{\xi}^{i}\gamma^{\mu}
\stackrel{\leftrightarrow}{\partial_{\mu}}\xi^{j}
+K^{i^{\ast}j}W^{\ast}_{i}W_{j} \nonumber\\
& &
+\frac{1}{2}iK_{pqi^{\ast}}\bar{\xi}^{i}\gamma^{\mu}\xi^{q}
\partial_{\mu}a^{p}
-\frac{1}{2}iK_{i^{\ast}p^{\ast}l}\bar{\xi}^{i}\gamma^{\mu}\xi^{l}
\partial_{\mu}a^{\ast p} \nonumber\\
& &
+\frac{1}{2}i\left( W_{ij} -K_{ijk^{\ast}}K^{k^{\ast}l}W_{l} \right)
\bar{\xi}^{ci}\xi^{j}
-\frac{1}{2}i\left( W^{\ast}_{ij}
-K_{i^{\ast}j^{\ast}k}K^{l^{\ast}k}W^{\ast}_{l} \right)
\bar{\xi}^{i}\xi^{cj} \nonumber\\
& &
+\frac{1}{4}\left( K_{ijk^{\ast}l^{\ast}}
-K^{m^{\ast}n}K_{ijm^{\ast}}K_{k^{\ast}l^{\ast}n} \right)
\bar{\xi}^{ci}\xi^{j}\bar{\xi}^{k}\xi^{cl}
-\frac{1}{4}\Box K(a^{i},a^{\ast j}),
  \label{eq:ord-2}
\end{eqnarray}
where $K$ is a K\"{a}hler potential and the lower indices $i$ of $K$
and $W$ mean the derivatives by $a^{i}$ and $i^{\ast}$means the
derivatives by $a^{\ast i}$,
\begin{eqnarray}
K_{i^{\ast}j} &=& \frac{\partial^{2}K(a,a^{\ast})}
{\partial a^{\ast i}\partial a^{j}},
  \label{eq:ord-3}\\
W_{i} &=& \frac{\partial W(a)}{\partial a^{i}}.
  \label{eq:ord-4}
\end{eqnarray}
$W^{\ast}_{i}$ means $(W_{i})^{\ast}$
and the orders of indices are commutable each other.
The matrix $K$ with upper
indices of $K$ is the inverse matrix of $K$:
\begin{eqnarray}
K^{ij^{\ast}}K_{j^{\ast}k}~=~\delta^{i}_{k} &,&~~
K_{i^{\ast}j}K^{jk^{\ast}}~=~\delta^{i^{\ast}}_{k^{\ast}}.
  \label{eq:ord-4a}
\end{eqnarray}

From this Lagrangian, canonical energy-momentum tensor $T^{\mu\nu}$
is given as follows:
\begin{eqnarray}
T^{00}&=&
\partial_{0}a^{\ast i}K_{i^{\ast}j}\partial_{0}a^{j}
+\partial_{1}a^{\ast i}K_{i^{\ast}j}\partial_{1}a^{j}
-\frac{1}{2}iK_{i^{\ast}j}\left(
\bar{\xi}^{i}\gamma^{1}\partial_{1}\xi^{j}
-\partial_{1}\bar{\xi}^{i}\gamma^{1}\xi^{j} \right) \nonumber\\
& &
-\frac{1}{2}iK_{i^{\ast}jk}\bar{\xi}^{i}\gamma^{1}\xi^{j}\partial_{1}a^{k}
+\frac{1}{2}iK_{ij^{\ast}k^{\ast}}\bar{\xi}^{j}\gamma^{1}\xi^{i}
\partial_{1}a^{\ast k}
+K^{i^{\ast}j}W^{\ast}_{i}W_{j} \nonumber\\
& &
-\frac{1}{2}i\left( W_{ij}-K^{km^{\ast}}K_{m^{\ast}ij}W_{k} \right)
\bar{\xi}^{ci}\xi^{j}
+\frac{1}{2}i\left(
W^{\ast}_{ij}-K^{k^{\ast}m}K_{mi^{\ast}j^{\ast}}W^{\ast}_{k} \right)
\bar{\xi}^{i}\xi^{cj} \nonumber\\
& &
-\frac{1}{4}\left( K_{i^{\ast}j^{\ast}kl}
-K^{mn^{\ast}}K_{i^{\ast}j^{\ast}m}K_{kln^{\ast}} \right)
\bar{\xi}^{i}\xi^{cj}\bar{\xi}^{ck}\xi^{l}
  \nonumber\\
T^{01}&=&
-\partial_{0}a^{\ast i}K_{i^{\ast}j}\partial_{1}a^{j}
-\partial_{1}a^{\ast i}K_{i^{\ast}j}\partial_{0}a^{j}
-\frac{1}{2}iK_{i^{\ast}j}\left(
\bar{\xi}^{i}\gamma^{0}\partial_{1}\xi^{j}
-\partial_{1}\bar{\xi}^{i}\gamma^{0}\xi^{j} \right) \nonumber\\
& &
-\frac{1}{2}iK_{i^{\ast}jk}\bar{\xi}^{i}\gamma^{0}\xi^{j}\partial_{1}a^{k}
+\frac{1}{2}iK_{ij^{\ast}k^{\ast}}\bar{\xi}^{j}\gamma^{0}\xi^{i}
\partial_{1}a^{\ast k}.
  \label{eq:ord-5}
\end{eqnarray}

The supercurrent $J^{\mu}$ is given by Noether procedure
\begin{eqnarray}
J^{\mu}&=&
\bar{\eta}^{c}j^{\mu} +\bar{j}^{\mu}\eta^{c}, \nonumber\\
j^{\mu}&=&
\sqrt{2}\left( \partial_{\nu}a^{\ast i}K_{i^{\ast}j}
\gamma^{\nu}\gamma^{\mu}\xi^{j}
-W^{\ast}_{i}\gamma^{\mu}\xi^{ci} \right), \nonumber\\
\bar{j}^{\mu}&=&
\sqrt{2}\left( \bar{\xi}^{i}\gamma^{\mu}\gamma^{\nu}K_{i^{\ast}j}
\partial_{\nu}a^{j}
-\bar{\xi}^{ci}\gamma^{\mu}W_{i} \right),
  \label{eq:ord-6}
\end{eqnarray}
where $\bar{\eta}^{c}$ and $\eta^{c}$ are the parameters of supersymmetry
transformation. The supercharge is defined by
\begin{eqnarray}
Q &=& \int_{-\infty}^{\infty} j^{0}(x) dx.
  \label{eq:ord-7}
\end{eqnarray}
In classical theory the orders of the operators appear in $j^{\mu}$ and
$\bar{j}^{\mu}$ are freely changed. But we have to fix the order of
the operators when we transfer from classical theory to quantum theory.
Our basic recipe to fix operator order is that
it gives the correct supersymmetry algebra.

Canonical momentum for $a$ and $\xi$ are given as follows:
\begin{eqnarray}
\pi_{a^{i}}&=&
\partial^{0}a^{\ast j}K_{j^{\ast}i}
+\frac{1}{2}iK_{ijk^{\ast}}\bar{\xi}^{k}\gamma^{0}\xi^{j}, \nonumber\\
\pi_{a^{\ast i}}&=&
K_{i^{\ast}j}\partial^{0}a^{j}
-\frac{1}{2}iK_{i^{\ast}j^{\ast}k}\bar{\xi}^{j}\gamma^{0}\xi^{k}, \nonumber\\
\pi_{\xi^{i}} &=&
\frac{1}{2}iK_{j^{\ast}i}\xi^{\dag j}, \nonumber\\
\pi_{\xi^{\dag i}}&=&
\frac{1}{2}iK_{i^{\ast}j}\xi^{j}.
  \label{eq:ord-8}
\end{eqnarray}

When the kinetic term has a flat K\"{a}hler potential,
$K(\phi,\phi^{\ast})=\phi^{\ast}\phi$, the supercurrent and
canonical momentum becomes as follows:
\begin{eqnarray}
j^{\mu}&=&
\sqrt{2}\left( \partial_{\nu}a^{\ast}\gamma^{\nu}\gamma^{\mu}\xi
-W'^{\ast}\gamma^{\mu}\xi^{c} \right),
  \label{eq:ord-8a}\\
\pi_{a}&=&
\partial^{0}a^{\ast},
  \label{eq:ord-8b}\\
\pi_{\xi} &=&
\frac{1}{2}i\xi^{\dag}.
  \label{eq:ord-8c}
\end{eqnarray}

In this case,
$K(\phi,\phi^{\ast})=\phi^{\ast}\phi$, there is no ordering problem in
$j^{\mu}$ because $\partial_{0}a^{\ast}$ and $\xi$ commute. Then we can
obtain the correct supersymmtry algebra \cite{mty}.

On the other hand,
when the kinetic term has a nonflat K\"{a}hler potential,
$K(\phi,\phi^{\ast}) \neq \phi^{\ast}\phi$,
$\partial_{0}a^{\ast}$ and $\xi$ become noncommutable.
Then we cannot change the order of bosons and fermions freely in $j^{\mu}$.
In the same way two orders in $\pi_{a^{i}}$,
$\partial^{0}a^{\ast j}K_{j^{\ast}i}$ and
$K_{j^{\ast}i}\partial^{0}a^{\ast j}$, are not equal.
So we have to use the correct operator order in
$j^{\mu}$ and $\pi_{a^{i}}$ to obtain the correct supersymmetry algebra.
As we will see later, the expressions
(\ref{eq:ord-6}) and (\ref{eq:ord-8}) give the
correct operator orders in $j^{\mu}$ and $\pi_{a^{i}}$, respectively.

On canonical quantization, canonical momenta for $\xi$ give primary
constraints:
\begin{eqnarray}
\chi_{\xi^{i}}&=&
\pi_{\xi^{i}} -\frac{1}{2}iK_{j^{\ast}i}\xi^{\dag j} ~=~0, \nonumber\\
\chi_{\xi^{\dag i}}&=&
\pi_{\xi^{\dag i}} -\frac{1}{2}iK_{i^{\ast}j}\xi^{j} ~=~0.
  \label{eq:ord-11}
\end{eqnarray}
These constraints are the second class constraints.

Canonical quantization condition is given through Dirac bracket.
There are 17 nonzero Dirac brackets in
55 Dirac brackets. There are 10 independent Dirac brackets in these
17 nonzero Dirac brackets. On calculation we use the operator order in
$\pi_{a}$ as (\ref{eq:ord-8}):
\begin{eqnarray}
\{ a^{i},\pi_{a^{j}} \}_{D} &=& \delta^{i}_{j}, \nonumber\\
\{ a^{\ast i},\pi_{a^{\ast j}} \}_{D} &=&
\delta^{i^{\ast}}_{j^{\ast}}, \nonumber\\
\{ \xi^{i},\xi^{\dag j} \}_{D} &=& -iK^{ij^{\ast}}, \nonumber\\
\{ \xi^{i},\pi_{a^{j}} \}_{D} &=&
-\frac{1}{2}K^{il^{\ast}}K_{l^{\ast}jm}\xi^{m}, \nonumber\\
\{ \xi^{i},\pi_{a^{\ast j}} \}_{D} &=&
-\frac{1}{2}K^{il^{\ast}}K_{l^{\ast}j^{\ast}m}\xi^{m}, \nonumber\\
\{ \xi^{\dag i},\pi_{a^{j}} \}_{D} &=&
-\frac{1}{2}K^{i^{\ast}l}K_{m^{\ast}jl}\xi^{\dag m}, \nonumber\\
\{ \xi^{\dag i},\pi_{a^{\ast j}} \}_{D} &=&
-\frac{1}{2}K^{i^{\ast}l}K_{m^{\ast}j^{\ast}l}\xi^{\dag m}, \nonumber\\
\{ \pi_{a^{i}},\pi_{a^{j}} \}_{D} &=&
-\frac{1}{4}iK^{kl^{\ast}}\left(
K_{m^{\ast}ik}K_{l^{\ast}jn} -K_{m^{\ast}jk}K_{l^{\ast}in} \right)
\xi^{\dag m}\xi^{n}, \nonumber\\
\{ \pi_{a^{i}},\pi_{a^{\ast j}} \}_{D} &=&
-\frac{1}{4}iK^{kl^{\ast}}\left( K_{m^{\ast}ik}K_{l^{\ast}j^{\ast}n}
-K_{m^{\ast}j^{\ast}k}K_{l^{\ast}in} \right)
\xi^{\dag m}\xi^{n}, \nonumber\\
\{ \pi_{a^{\ast i}},\pi_{a^{\ast j}} \}_{D} &=&
-\frac{1}{4}iK^{kl^{\ast}}\left( K_{m^{\ast}i^{\ast}k}K_{l^{\ast}j^{\ast}n}
-K_{m^{\ast}j^{\ast}k}K_{l^{\ast}i^{\ast}n} \right)
\xi^{\dag m}\xi^{n}.
  \label{eq:ord-12}
\end{eqnarray}
Replacing these Dirac brackets with (anti)commutator divided by $i$,
we obtain the following canonical quantization conditions:
\begin{eqnarray}
\lbrack a^{i}(x,t), \partial_{0}a^{\ast j}(y,t) \rbrack &=&
K^{ij^{\ast}}(x,t)
\cdot i\delta(x-y), \nonumber\\
\lbrack a^{\ast i}(x,t), \partial_{0}a^{j}(y,t) \rbrack &=&
K^{i^{\ast}j}(x,t)
\cdot i\delta(x-y), \nonumber\\
\{ \xi^{i}(x,t), \xi^{\dag j}(y,t) \} &=&
-iK^{ij^{\ast}}(x,t)I
\cdot i\delta(x-y), \nonumber\\
\lbrack \xi^{i}(x,t), \partial_{0}a^{\ast j}(y,t) \rbrack &=&
-\left( K^{im^{\ast}}K^{j^{\ast}n}K_{m^{\ast}nl}\xi^{l} \right)
(x,t)\cdot i\delta(x-y), \nonumber\\
\lbrack \xi^{\dag i}(x,t), \partial_{0}a^{j}(y,t) \rbrack &=&
-\left( K^{i^{\ast}m}K^{jn^{\ast}}K_{mn^{\ast}l^{\ast}}\xi^{\dag l} \right)
(x,t)
\cdot i\delta(x-y), \nonumber\\
\lbrack \partial_{0}a^{\ast i}(x,t),
\partial_{0}a^{j}(y,t) \rbrack &=&
\left( -K^{jl^{\ast}}\partial_{0}a^{\ast m}K_{m^{\ast}l^{\ast}k}K^{ki^{\ast}}
+K^{jl^{\ast}}K^{ki^{\ast}}K_{mkl^{\ast}}\partial_{0}a^{m} \right. \nonumber\\
& &
+iK^{jl^{\ast}}K^{ki^{\ast}}K^{pq^{\ast}}K_{nkq^{\ast}}K_{m^{\ast}l^{\ast}p}
\xi^{\dag m}\xi^{n} \nonumber\\
& &
\left.
-iK^{jl^{\ast}}K^{ki^{\ast}}K_{knm^{\ast}l^{\ast}}\xi^{\dag m}\xi^{n}
\right)
(x,t)\cdot i\delta(x-y).
  \label{eq:ord-13}
\end{eqnarray}
\section{Fixing of operator order and SUSY algebra}

We have to fix the order of the operators appear in $j^{\mu}$
to transfer from classical theory to quantum theory.
Our basis to obtain correct operator order is that
it gives the correct supersymmetry algebra. We select
the operator order which satisfy the following relation:
\begin{eqnarray}
-i\lbrack \varphi, Q \rbrack_{\pm}
-i\lbrack \varphi, \bar{Q} \rbrack_{\pm} &=&
\delta\varphi,
  \label{eq:ord-14}
\end{eqnarray}
where $\varphi$ represents each component fields
\begin{eqnarray}
\delta a^{i} &=&
\sqrt{2}\bar{\eta}^{c}\xi^{i} \nonumber\\
\delta \xi^{i} &=&
-\sqrt{2}i\partial_{\mu}a^{i}\gamma^{\mu}\eta^{c}+\sqrt{2}\eta f^{i}
  \label{eq:ord-15}
\end{eqnarray}
and where $f^{i}$ are auxiliary fields which satisfy
\begin{eqnarray}
f^{i} &=&
\frac{1}{2}K^{im^{\ast}}K_{m^{\ast}kl}\bar{\xi}^{cl}\xi^{k}
-iK^{ij^{\ast}}W^{\ast}_{j}.
  \label{eq:ord-16}
\end{eqnarray}
The supercurrent which satisfies the above relation (\ref{eq:ord-14})
gives the correct supersymmetry algebra. 

With the supercurrent in the previous operator order (\ref{eq:ord-6}),
each component fields satisfy these relations (\ref{eq:ord-14}).
Then the supercurrent in this operator order gives a correct operator
order because it gives the correct supersymmetry algebra as follows:
\begin{eqnarray}
\{ Q,\bar{Q} \} ~=~ 2\gamma^{\mu}P_{\mu},~~
\{ Q,\bar{Q}^{c} \} ~=~ T\gamma_{5}
  \label{eq:ord-18}
\end{eqnarray}
provided that the operator order in $\pi_{a^{i}}$ is given as
(\ref{eq:ord-8}). The other supercurrents which are equivalent to
(\ref{eq:ord-6}) also satisfy (\ref{eq:ord-14}) and then they are also
correct operator orders. But the supercurrents which are not equivalent to
(\ref{eq:ord-6}) do not satisfy (\ref{eq:ord-14}) and then they do not
give the correct supersymmerty algebra. So they are wrong operator orders.
Then the relation (\ref{eq:ord-14}) plays a role to select the correct
operator orders from all possible orders. 

In the above, we use the operator order (\ref{eq:ord-8}) as $\pi_{a^{i}}$.
The other order of $\pi_{a^{i}}$, $K_{j^{\ast}i}\partial^{0}a^{\ast j}$,
does not give the supersymmetry algebra correctly.
Then the operator order (\ref{eq:ord-8}) gives the correct operator
order of $\pi_{a^{i}}$.

In the above supersymmtry algebra $T$ is given as follows:
\begin{eqnarray}
T &=&
-4\int_{-\infty}^{\infty} \partial_{1}\left( W^{\ast}(a^{\ast}(x)) \right)dx
\nonumber\\
&=&
-4\left\{ W^{\ast}(a^{\ast}(x=\infty)) - W^{\ast}(a^{\ast}(x=-\infty))
\right\}
\nonumber\\
&\equiv&
-4\Delta W^{\ast}.
  \label{eq:ord-19}
\end{eqnarray}
If the superpotential $W^{\ast}(a^{\ast}(x))$
have different values at
$x=\infty$ and $x=-\infty$, solitonic configurations of $a(x)$,
$T$ has a nonzero value.
In this case it gives a central extension of supersymmetry algebra.
\section{Classical Soliton path in complex $W$-plane}

We assume that $W(\phi^{i})$ is a holomorphic function such that
$W_{i}=0$ has $N$ complex solutions
\begin{eqnarray}
a_{(I)} &=&
\left ( a_{(I)}^{1},a_{(I)}^{2},\cdots,a_{(I)}^{n} \right)
  \label{eq:hj-1}
\end{eqnarray}
where $I$ runs from 1 to $N$.

There are not only $N$ classical vacuum configurations
$a(t,x)=a_{(I)}$
but also solitonic configurations.
There are at most $N(N-1)$ solitonic configurations characterized by
\begin{eqnarray}
a(t,-\infty) ~=~ a_{(I)},~~
a(t,\infty) ~=~ a_{(J)}
  \label{eq:hj-2}
\end{eqnarray}
which we call $(I,J)$-soliton.

In the presence of $(I,J)$-soliton, supersymmetry algebra has a central
charge (\ref{eq:ord-18}). In the center of mass frame
$(P^{\mu})=(M_{IJ},0)$ it becomes
\begin{eqnarray}
\{ Q_{\alpha},Q^{\dag}_{\beta} \} ~=~
2M_{IJ}\delta_{\alpha\beta},~~
\{ Q_{\alpha},Q_{\beta} \} ~=~
-4i(\sigma_{1})_{\alpha\beta}\Delta_{IJ}W^{\ast}
  \label{eq:hj-2a}
\end{eqnarray}
where $\Delta_{IJ}W=W(a_{(J)})-W(a_{(I)})$ and $\alpha, \beta$ run 1 and 2.
We put
\begin{eqnarray}
A ~=~ Q_{1}+ie^{i\theta}Q^{\dag}_{2},~~
B ~=~ Q_{1}-ie^{i\theta}Q^{\dag}_{2}
  \label{eq:hj-2b}
\end{eqnarray}
then we have Bogomol'nyi bound of $(I,J)$-soliton 
\begin{eqnarray}
M_{IJ} \geq 2|\Delta_{IJ}W|
  \label{eq:hj-2c}
\end{eqnarray}
from positivity of $A$
\begin{eqnarray}
\{ A,A^{\dag} \} \geq 0.
  \label{eq:hj-2d}
\end{eqnarray}

Applying Hamilton-Jacobi method of classical mechanics,
we can show that this Bogomol'nyi bound is saturated by classical solution.
Dropping the fermion degrees of freedom from $T^{00}$ and considering
the static configuration, and regarding $x$ as time, we have the following
``Lagrangian'':
\begin{eqnarray}
{\cal L}'&=&
\partial_{1}a^{\ast i}K_{i^{\ast}j}\partial_{1}a^{j}
+K^{i^{\ast}j}W^{\ast}_{i}W_{j}.
  \label{eq:hj-3}
\end{eqnarray}
From this ``Lagrangian'', we have the following Hamiltonian:
\begin{eqnarray}
{\cal H}'&=&
K^{kl^{\ast}}\left(
p_{a^{k}}p_{a^{\ast l}} -W_{k}W^{\ast}_{l} \right),
  \label{eq:hj-4}
\end{eqnarray}
where $p_{a^{k}}$ is a conjugate momentum to $a^{k}$.
This is a problem of a classical particle moving in the potential
\begin{eqnarray}
U&=&-K^{i^{\ast}j}W^{\ast}_{i}W_{j}.
  \label{eq:hj-5}
\end{eqnarray}
We apply Hamilton-Jacobi method to this problem.
The Hamilton-Jacobi equation for the action $S(a^{i},a^{\ast i})$ is
\begin{eqnarray}
K^{ij^{\ast}}\left(
\frac{\partial S}{\partial a^{i}} \frac{\partial S}{\partial a^{\ast j}}
-W_{i}W^{\ast}_{j} \right)
&=&E.
  \label{eq:hj-6}
\end{eqnarray}
The following action solves 
the equation (\ref{eq:hj-6}) for $E=0$
\begin{eqnarray}
S(a^{i},a^{\ast i},\alpha)&=&
\alpha W(a^{i}) +\frac{1}{\alpha}W^{\ast}(a^{\ast i}),
  \label{eq:hj-7}
\end{eqnarray}
where $\alpha=e^{i\omega}$ is a parameter. Except $n=1$ case,
this action is not a perfect solution because it has only one parameter.
So it is a nonperfect solution of Hamilton-Jacobi equation.
Then it does not fix a path which connect two classical vacua in
complex $n$-dimensional $a^{i}$-space. When $n=1$ case it becomes a perfect
solution and then it describes a path which connects two classical vacua
in complex one-dimensional $a$-plane.

But we can find a curve of soliton path which connects two classical
vacua in complex $W$-plane as a projection of the soliton path in
complex n-dimensional $a^{i}$-space.
The action $S$ satisfies the following condition:
\begin{eqnarray}
\frac{\partial S}{\partial\alpha} &=& const.
  \label{eq:hj-8}
\end{eqnarray}
Then we have
\begin{eqnarray}
{\rm Im}\left(e^{i\omega}W(a^{i})\right) &=& const.
  \label{eq:hj-9}
\end{eqnarray}
It means that we have a straight line of soliton path in
complex $W$-plane although
in general soliton path is a curve in complex $n$-dimensional $a^{i}$-space.
When $n=1$ we can find a path in complex 1-dimensional $a$-plane by
solving (\ref{eq:hj-9}) to $a$.

Around $a^{i}=a_{(I)}^{i}$
superpotential $W(a)$ is expanded by power series of
($a^{i}-a_{(I)}^{i}$):
\begin{eqnarray}
W(a) &=&
W(a_{(I)})
+\frac{1}{2}W_{ij}(a_{(I)})(a^{i}(x)-a_{(I)}^{i})(a^{j}(x)-a_{(I)}^{j})
\nonumber\\
& &
+\frac{1}{3!}W_{ijk}(a_{(I)})(a^{i}(x)-a_{(I)}^{i})
(a^{j}(x)-a_{(I)}^{j})(a^{k}(x)-a_{(I)}^{k})
+\cdots.
  \label{eq:hj-10}
\end{eqnarray}
$W(a)$ is not a single valued function around $W(a_{I})$.
Then there is a branch cut extending from each $W(a_{I})$ to infinity.
So only soliton paths which do not cross branch cuts can exist.

$S$ also satisfies
\begin{eqnarray}
p_{a^{k}} &=&
\frac{\partial S}{\partial a^{i}}.
  \label{eq:hj-10a}
\end{eqnarray}
Then we have
\begin{eqnarray}
K_{i^{\ast}j}\partial_{1}a^{\ast i}\partial_{1}a^{j}
&=&
\frac{|dW|}{dx}
  \label{eq:hj-10b}
\end{eqnarray}
because $dW = |dW|\alpha^{-1}$.

Classical mass of $(I,J)$-soliton is given by
\begin{eqnarray}
M_{IJ}&=&
\int_{-\infty}^{\infty} {\cal L}' dx
~=~
2\int_{-\infty}^{\infty}
K_{i^{\ast}j}\partial_{1}a^{\ast i}\partial_{1}a^{j} dx \nonumber\\
&=&
2\int_{W(a_{I})}^{W(a_{J})} |dW|
~=~
2\left| \int_{W(a_{I})}^{W(a_{J})} dW \right|
~=~
2|\Delta_{IJ} W|
  \label{eq:hj-11}
\end{eqnarray}
since $W(a)$ is a straight line in the complex $W$-plane.
This shows the Bogomol'nyi bound is saturated by classical solution.
Classical mass of $(I,J)$-soliton is given by the length of its path
in the complex $W$-plane. Then we have a inequality among the masses
\begin{eqnarray}
M_{IK} < M_{IJ}+M_{JK}.
  \label{eq:hj-12}
\end{eqnarray}
The equality $M_{IK}=M_{IJ}+M_{JK}$ does not appear because in this case
$(I,K)$-soliton path
crosses $a_{(J)}$ between $a_{(I)}$ and $a_{(K)}$ and
it takes twice of infinite ``time''. It means that there is attractive
force between neighboring solitons.
\section{Multiplet shortening of supersymmetry Algebra}

In the presence of $(I,J)$-soliton and static case, each components of
supercharge are given as
\begin{eqnarray}
Q_{1}&=&
\sqrt{2}\int \left(
\partial_{1}a^{\ast i}K_{i^{\ast}j}\xi^{j}_{1}+iW^{\ast}_{i}\xi^{\ast i}_{2}
\right) dx, \nonumber\\
Q_{2}&=&
\sqrt{2}\int \left(
-\partial_{1}a^{\ast i}K_{i^{\ast}j}\xi^{j}_{2}-iW^{\ast}_{i}\xi^{\ast i}_{1}
\right) dx.
  \label{eq:ms-1}
\end{eqnarray}
From these expressions $A$ and $B$ are given as
\begin{eqnarray}
A&=&0, \nonumber\\
B&=&
\sqrt{2}\int \left(
e^{i\theta}W_{i}\xi^{i}_{1}+iW^{\ast}_{i}\xi^{\ast i}_{2} \right) dx.
  \label{eq:ms-2}
\end{eqnarray}
Then $\{ A,A^{\dag} \}=0$ means the Bogomol'nyi bound is saturated.
And the anticommutator for $B$ becomes
\begin{eqnarray}
\{ B,B^{\dag} \} &=& 8M_{IJ}.
  \label{eq:ms-3}
\end{eqnarray}
So only $B$ and $B^{\dag}$ can excite the fermionic modes.
Then the number of bosonic and fermionic state is a half of that without
$(I,J)$-soliton. If there is no $(I,J)$-soliton, then we have
no central charge, so both $A$, $A^{\dag}$ and $B$, $B^{\dag}$ excite
the fermionic modes.
Then a half of supersymmetry is broken by the existence of $(I,J)$-soliton.
\section{Commutation relations for $T$}

Whether the operator $T$ commutes with the other operators or not is not
trivial. We check that the operator $T$ commutes with the other operators.

Obviously $T$ commutes with itself,
\begin{eqnarray}
[ T,T ] &=& 0.
  \label{eq:cr-1}
\end{eqnarray}
$T$ also commutes with $Q$ because $Q$ do not contain $\partial_{0}a^{i}$,
\begin{eqnarray}
[ T,Q ] &=& 0.
  \label{eq:cr-2}
\end{eqnarray}
$T$ commutes with $\bar{Q}$, $P^{0}$ and $P^{1}$,
\begin{eqnarray}
\lbrack T,\bar{Q} \rbrack ~=~
-4\sqrt{2}i \int \partial_{1}\left( \bar{\xi}^{i}W^{\ast}_{i} \right) dx
~=~ 0,
  \label{eq:cr-3}\\
\lbrack T,P^{0} \rbrack ~=~
-4i \int \partial_{1}\left( \partial_{0}a^{\ast i}W^{\ast}_{i} \right) dx
~=~ 0,
  \label{eq:cr-4}\\
\lbrack T,P^{1} \rbrack ~=~
-4i \int \partial_{1}\left( \partial_{1}a^{\ast i}W^{\ast}_{i} \right) dx
~=~ 0.
  \label{eq:cr-5}
\end{eqnarray}

We calculate commutation relation for $T$ and the angular momentum operator.
In 2-dimension the angular momentum operator $M^{01}$ is given as
\begin{eqnarray}
M^{01} &=&
\int \left( T^{00}x -T^{01}t \right) dx.
  \label{eq:cr-6}
\end{eqnarray}
For the $T^{01}$ part
\begin{eqnarray}
[ T, \int T^{01}(y)t dy ] ~=~
-4i \int \partial_{1}\left( \partial_{1}a^{\ast i}W^{\ast}_{i}t \right) dx
~=~ 0.
  \label{eq:cr-7}
\end{eqnarray}
For the  $T^{00}$ part
\begin{eqnarray}
[ T, \int T^{00}(y)y dy ] &=&
-4i \int \partial_{1}\left( \partial_{0}a^{\ast i}W^{\ast}_{i}x \right) dx.
  \label{eq:cr-8}
\end{eqnarray}
If $W^{\ast}_{i}$ drops faster than $x^{-1}$ then this commutator vanishes.
From (\ref{eq:hj-10a}) we have
\begin{eqnarray}
K_{i^{\ast}j}\partial_{1}a^{\ast i} &=& W_{j}\alpha.
  \label{eq:cr-9}
\end{eqnarray}
$W_{j}(a)$ is expanded by power series around $a^{i}=a_{(I)}^{i}$:
\begin{eqnarray}
W_{j}(a) &=&
W_{jl}(a_{(I)})(a^{l}(x)-a_{(I)}^{l}) +\cdots.
  \label{eq:cr-10}
\end{eqnarray}
Then (\ref{eq:cr-9}) becomes
\begin{eqnarray}
K_{i^{\ast}j}(a_{(I)})\frac{\partial a^{\ast i}}{\partial x} &=&
\alpha W_{ij}(a_{(I)})(a^{i}(x)-a_{(I)}^{i})
  \label{eq:cr-11}
\end{eqnarray}
around $a^{i}=a_{(I)}^{i}$.
In general a soliton path is a curve in the complex $a$-plane, but
in the small area around $a^{i}=a_{(I)}^{i}$ the soliton path is
approximated by a straight line. We rotate the real axis along this
short line, we can take $a^{i}$ as real near $a^{i}=a_{(I)}^{i}$:
\begin{eqnarray}
K_{ij}(a_{(I)})\frac{\partial a^{i}}{\partial x} &=&
\alpha W_{ij}(a_{(I)})(a^{i}(x)-a_{(I)}^{i}).
  \label{eq:cr-12}
\end{eqnarray}
Since $K_{ij}$ and $W_{ij}$ are complex symmetric matrices,
we can diagonalize them by unitary matrix. Then $a(x)$ has a form
\begin{eqnarray}
K_{ii}(a_{(I)})\frac{\partial a^{i}}{\partial x} &=&
\alpha W_{ii}(a_{(I)})(a^{i}(x)-a_{(I)}^{i}),
  \label{eq:cr-13}
\end{eqnarray}
where $K_{ij}$ and $\alpha W_{ij}$ must have same phase because
$\partial a^{i} / \partial x$ and $a^{i}(x)-a_{(I)}^{i}$ are real.
We drop this phase, we have
\begin{eqnarray}
\frac{\partial a^{i}}{\partial x} &=&
const. \times (a^{i}(x)-a_{(I)}^{i}).
  \label{eq:cr-14}
\end{eqnarray}
This equation has a solution
\begin{eqnarray}
a^{i}(x)-a_{(I)}^{i} &=&
(a^{i}(x_{0})-a_{(I)}^{i}) e^{const. \times (x-x_{0})}.
  \label{eq:cr-15}
\end{eqnarray}
We can take the sign of the exponent properly.
Then $a^{i}$ converges exponentially, $W_{i}(a)$ also converges
exponentially, and (\ref{eq:cr-8}) becomes zero.
So the operator $T$ commutes with all the other operators, therefore
we can call it as the central charge.
\section{Conclusion}

We investigated a two-dimensional $N=2$ supersymmetric model which
consists of $n$ chiral superfields with K\"{a}hler potential.

In the first half, we argued about the operator ordering problem.
When the kinetic term has a flat K\"{a}hler potential,
$K(\phi ,\phi^{\ast})=\phi\phi^{\ast}$, there is no ordering problem.
On the other hand, when the kinetic term has a nonflat K\"{a}hler
potential $K(\phi ,\phi^{\ast}) \neq \phi\phi^{\ast}$, there arise
ordering problem. We saw that general K\"{a}hler potential raises
the ordering problem. When we transfer from classical theory to
quantum theory, there are several ways to fix the operator orders.
In the presence of supersymmetry it dictates
the operator ordering: we can fix the operator orders by requiring
the super-Poincar\'{e} algebra. It is also true when the supersymmetry
algebra has a central extension by the presence of topological soliton.

In the latter half, we argued about some natures of soliton in this model.
In general the path of soliton is a curve in complex $n$-dimensional
$a^{i}$-space and it is difficult to find it except $n=1$ case.
But Hamilton-Jacobi method of classical mechanics leads the result that
in complex $W$-plane we can find the path of the soliton and which is a
projection of the path in complex $n$-dimensional $a^{i}$-space.
The path of the soliton is a straight line in complex $W$-plane.
Classical mass of the soliton is given by the length of its path
in complex $W$-plane. Then we obtain a triangular inequality among the
masses. It means that there is attractive force between neighboring solitons.
In the presence of soliton, a half of supersymmetry is broken
because the Bogomol'nyi bound is saturated.
\section*{Acknowledgments}

The author (M. Y.) would like to thank K. Nishio and
the auther (N. M.) would like to thank S. Onizawa for useful
discussions.

\end{document}